# Deludedly Agreeing to Agree

Ziv Hellman[*]
Department of Statistics and Operations Research
The School of Mathematical Sciences
Tel Aviv University
Tel Aviv, 69978, Israel
zivhellman@post.tau.ac.il

## ABSTRACT

We study conditions relating to the impossibility of agreeing to disagree in models of interactive KD45 belief (in contrast to models of S5 knowledge, which are used in nearly all the agreements literature). We show that even when the truth axiom is not assumed it turns out that players will find it impossible to agree to disagree under fairly broad conditions.[1]

## General Terms

Theory

## 1. INTRODUCTION

One of the strongest assumptions underpinning the standard model of knowledge, known as S5, is the *truth axiom*, which essentially states that 'everything that a player knows is true'. This is equivalent, from one perspective, to asserting that no mistakes are ever made in the processing of signals.

Mistakes, of course, abound around us, and sometimes such mistakes can have significant consequences. Consider, for example the following scenario (a variation of an example appearing in [Hart and Tauman (2004)]): There are two traders. They trade on a daily basis, and since a trade involves one trader selling and the other buying, they can at least observe each others' willingness to trade. We may imagine that these two traders are the 'market leaders', in the sense that their actions are followed by others in the market and copied.

Let $\Omega$ be the set of all states of the world, with $\Omega$ containing nine states; $\Omega = \{1, 2, \ldots, 9\}$. For simplicity we will assume that there is a common prior $p$ over $\Omega$, with $p(\omega) = 1/9$ for all states $\omega$. The private information of the two traders, Anne and Bob are summarized by partitions $\Pi_A$ and $\Pi_B$ respectively, with

$$\Pi_A = 1234|5678|9$$

and

$$\Pi_B = 123|456|789.$$

One standard interpretation of the structure of such partitional knowledge is that Anne and Bob receive signals. If the true state is 2, for example, Anne receives a signal that enables her to rule out the states $5, 6, 7, 8, 9$, and she therefore knows that the true state is one of $1, 2, 3, 4$. Bob, at the true state 3, receives a signal that enables him to rule out the states $4, 5, 6, 7, 8, 9$, and he therefore knows that the true state is one of $1, 2, 3$. Specifically, suppose that Bob may receive any one of three signals, $\sigma_1, \sigma_2, \sigma_3$, where $\sigma_1$ informs Bob that the true state is one of $1, 2, 3$, $\sigma_2$ informs Bob that the true state is one of $4, 5, 6$, and $\sigma_3$ informs Bob that the true state is one of $7, 8, 9$ (we will be less interested in this example with specifying Anne's possible signals).

| Signal | States |
|---|---|
| $\sigma_1$ | $\rightarrow \{1, 2, 3\}$ |
| $\sigma_2$ | $\rightarrow \{4, 5, 6\}$ |
| $\sigma_3$ | $\rightarrow \{7, 8, 9\}$ |

Figure 1: Bob's signals and their interpretation when there are no processing errors.

So far, so standard. Now consider the possibility of a mistake in signals processing on the part of Bob. Suppose that Bob inputs the signals he receives into a black box that he has been assured outputs $1, 2, 3$, $4, 5, 6$, or $7, 8, 9$ if the input is $\sigma_1, \sigma_2$, or $\sigma_3$ respectively. Unbeknownst to Bob (and to Anne), however, Bob's black box is defective; when either $\sigma_1$ or $\sigma_2$ is given as input, the box outputs $4, 5, 6$ (hence even if, e.g., the true state is 1 Bob thinks the true state is one of $\{4, 5, 6\}$).

| Signal | States |
|---|---|
| $\sigma_1$ | $\rightarrow \{4, 5, 6\}$ |
| $\sigma_2$ | $\rightarrow \{4, 5, 6\}$ |
| $\sigma_3$ | $\rightarrow \{7, 8, 9\}$ |

Figure 2: Bob's signal processing error.

Consider next the event $E = \{4, 9\}$. This event will be interpreted as a 'good' outcome (e.g. company earnings are about to rise), with the complement representing a 'bad' event that ought to trigger the sale of shares. Suppose that

---

[*]Research supported in part by the European Research Council under the European Commission's Seventh Framework Programme (FP7/2007 - 2013)/ERC grant agreement no. 249159, and in part by Israel Science Foundation grants 538/11 and 212/09.

[1] What follows is an extended abstract for TARK, not a full paper.

the true state is 2, and that each one of the two traders behaves each day according to the following rule:

$$\begin{cases} \text{Buy} & \text{if the probability of } E \text{ is 0.3 or more;} \\ \text{Sell} & \text{if the probability of } E \text{ is less than 0.3.} \end{cases}$$

Given these assumptions, the following sequence of actions transpires. On Day 1, Anne, who processes signals correctly, supposes that the true state is one of $1, 2, 3, 4$, judges the probability of $E$ to be $1/4$ and seeks to sell shares. Bob erroneously supposes that the true state is one of $4, 5, 6$, judges the probability of $E$ to be $1/3$, and therefore buys shares from Anne.

Since Bob was willing to buy on Day 1, Anne 'learns' that the true state is not in $1, 2, 3$. She therefore erroneously supposes on Day 2 that the true state is 4 and offers to buy on Day 2. Bob does the same. By Day 3, it is 'common knowledge' that 4 is the 'true state' – Bob's error has now become Anne's error. Both traders seek to buy as many shares as they can, to their detriment, and a bubble has developed.

[Geanakoplos (1989)] and [Morris (1996)] show that in knowledge models that satisfy the truth axiom (but are not necessarily S5) more information is always beneficial for a player, in the sense that with more information a rational player will never choose an action that gives him less in expectation than an action that he chooses when he has less information. Without the truth axiom, that no longer holds true. Indeed, as the example here shows, without the truth axiom, not only is the 'mistaken' player in danger of choosing detrimental actions, his errors can cascade and 'infect' other players to their detriment: in Day 1 above, Anne makes the right decision in seeking to sell shares, but on Day 2, due to Bob's mistake, she is buying shares. Arguably, Anne has been mistaken all along, in accepting Bob's reports at face value, without considering the possibility that Bob might be mistaken.

The above story motivates the study of agreement and disagreement in models of *belief* as opposed to models of *knowledge*, which is the standard setting of most of the agreement literature.

## 2. PRELIMINARIES

### 2.1 Belief Structures

Fix a finite set of players $I$ and a finite set of *states of the world*[2] denoted by $\Omega$. Subsets of $\Omega$ are called *events*. The set of probability distributions over $\Omega$ is denoted by $\Delta(\Omega)$.

A *type function* $t_i$ over $\Omega$ for player $i$ is defined by assigning, for each $\omega$, a probability distribution $t_i(\omega) \in \Delta(\Omega)$ representing player $i$'s beliefs at $\omega$. We associate with each type function $t_i$ a partition $\Pi_i$ of $\Omega$ defined[3] by $\Pi_i(\omega) = \{\omega' \mid t_i(\omega') = t_i(\omega)\}$. If we impose on a type function the property that $t_i(\omega)(\Pi_i(\omega)) = 1$, then the type functions is *partitional*. A *probabilistic belief structure* over $\Omega$ is then a set of partitional type functions $(t_i)_{i \in I}$ over $\Omega$.

A function $b_i : \Omega \to 2^\Omega \setminus \emptyset$ is a *possibility function*. The event $b_i(\omega)$ is interpreted as the set of states that are considered possible for i at $\omega$, while all other states are excluded by i at $\omega$. We will call a possibility function $b_i : \Omega \to 2^\Omega \setminus \{\emptyset\}$ that is measurable with respect to a partition $\Pi_i$ and satisfies $b_i(\omega) \subseteq \Pi_i(\omega)$ for each $\omega \in \Omega$ a *KD45 possibility function*.

A *belief structure* over $\Omega$ is a set of pairs $\mathbf{\Pi} = (\Pi_i, b_i)_{i \in I}$, where each $b_i$ is a KD45 possibility function with respect to the partition $\Pi_i$ of $\Omega$. We will sometimes also call such a structure a KD45 belief structure.

The general structure of a model of KD45 belief of a player $i$ is of an over-arching partition $\Pi_i$, with each partition element $\pi \in \Pi_i$ furthermore partitioned into $b_i(\omega)$ and $f_i(\omega)$ (using an arbitrary $\omega \in \pi$). Every element $\omega' \in f_i(\omega)$ is mapped by $b_i$ into $b_i(\omega)$, where it is 'trapped', in the sense that $b_i(b_i(\omega')) = b_i(\omega') = b_i(\omega)$.

A probabilistic belief structure $(t_i)_{i \in I}$ over $\Omega$ induces a belief structure $(\Pi_i, b_i)_{i \in I}$ over $\Omega$, where $\Pi_i$ is the partition of $\Omega$ into the types of player $i$ and $b_i(\omega)$ is the set of states in $\Pi_i(\omega)$ that have positive $t_i(\omega)$ probability. Conversely, every belief structure over $\Omega$ is induced by a probabilistic belief structure over $\Omega$. We will sometimes make use of this by choosing, for a given belief structure $\mathbf{\Pi} = (\Pi_i, b_i)_{i \in I}$, an arbitrary probabilistic belief structure $(t_i^b)_{i \in I}$ that induces $\mathbf{\Pi}$.

### 2.2 Delusion

Let $\mathbf{\Pi} = (\Pi_i, b_i)_{i \in I}$ be a belief structure. If $\omega \in b_i(\omega)$ then $b_i$ is *non-deluded* at $\omega$. If $\omega \notin b_i(\omega)$ then $b_i$ is *deluded* at $\omega$; in this case we will also sometimes say that $\omega$ is a deluded state for player $i$. This is where we are dropping the 'truth axiom': if one accepts the truth axiom there are never any deluded states for any player.

If there is at least one state at which $b_i$ is deluded, then $b_i$ is *delusional*, and we will similarly say that the corresponding belief operator $B_i$ is delusional if this is the case. It is straight-forward to show that a belief structure $\mathbf{\Pi}$ is non-delusional for all players if and only if it is an S5 structure, and it is similarly straight-forward to show that a state $\omega$ is non-deluded for player $i$ if and only if $t_i^b(\omega) = 0$ for any probabilistic belief structure $(t_i^b)_{i \in I}$ that induces $\mathbf{\Pi}$.

*Definition 1.* A KD45 belief structure at which at all states $\omega \in \Omega$ either a) every player $i$ is deluded at $\omega$ or b) every player $i$ is non-deluded at $\omega$ will be called a *non-singular* structure. ♦

In examples, we will compactly express KD45 belief structures by separating states in different partition elements of $\Pi_i$ by the square boxes. Within each partition element we will denote states that are in the same component of $b_i(\omega)$ by an oval box.

For example, if we write

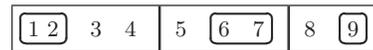

then the intention is, for example, that 5, 6 and 7 are all in the same partition element, i.e., $\Pi_i(5) = \{5, 6, 7\}$, but 5 is a delusional state such that $b_i(5) = \{6, 7\}$.

## 3. BELIEF REVISION

The general approach we will follow is: in standard S5 concepts and formulae, replace $\Pi_i$ by $b_i$ and see what happens. We will apply this now to Bayesian belief revision.

---

[2] In the basic definitions of elements of belief structures we largely follow [Samet (2011)].

[3] The presentation here reverses most presentations of belief structures, in which partitions are given and used to define type functions; here we are starting with type functions and using them to define the partitions.

## 3.1 Standard belief revision and priors

Let $\mu$ be a probability distribution over $\Omega$, and let $\Pi_i$ be a partition of of $\Omega$. The *(standard) revision* of $\mu$ at $\omega$ according to $\Pi_i$ is the probability distribution $\widehat{\mu}(\omega)$ such that

$$\widehat{\mu}(\omega)(\omega') = \begin{cases} \frac{\mu(\omega')}{\mu(\Pi_i(\omega))} & \text{if } \omega' \in \Pi_i(\omega) \\ 0 & \text{otherwise} \end{cases} \quad (1)$$

if $\mu(\Pi_i(\omega)) > 0$; otherwise it is undefined.

We may interpret this as follows: ex ante the player has a prior probability distribution of full support. When updating following a signal, the player excludes states outside $b_i(\omega)$, i.e. gives them zero probability. Since the player mistakes the reading of the signal, it is possible that he or she ends up giving the true state $\omega$ zero probability.

Let $f$ be a random variable over $\Omega$, $\mu$ be a probability distribution over $\Omega$, and $\Pi_i$ a partition of $\Omega$. Then the *conditional expected value of $f$ at $\omega$* is

$$E_i^\mu(f \mid \Pi_i(\omega)) := \frac{1}{\mu(\Pi_i(\omega))} \sum_{\omega' \in \Pi_i(\omega)} f(\omega')\mu(\omega)(\omega'), \quad (2)$$

if $\mu(\Pi_i(\omega)) \neq 0$ (otherwise it is not defined).

Let $(t_i)_{i \in I}$ be a probabilistic belief structure over $\Omega$, with $(\Pi_i)_{i \in I}$ the corresponding partition. A *(standard) prior* for $t_i$ is a probability distribution $\mu \in \Delta(\Omega)$, such that $\widehat{\mu}(\omega) = t_i(\omega)$ at each $\omega$, where $\widehat{\mu}(\omega)$ is the standard revision of $\mu$ at $\omega$ according to $\Pi_i$ as defined in Equation (1). A *(standard) common prior* for $(t_i)_{i \in I}$ is a probability distribution $\mu \in \Delta(\Omega)$ that is a prior for each $t_i$.

Given a probabilistic belief structure $(t_i)_{i \in I}$ with corresponding partition $(P_i)_{i \in I}$, player $i$'s *posterior expected value of $f$ at $\omega$* is

$$E_i^{t_i}(f \mid \Pi_i(\omega)) := \sum_{\omega' \in \Pi_i(\omega)} t_i(\omega')f(\omega'). \quad (3)$$

If there is a common prior $\mu$, then for any random variable $f$ the posterior expected value of each player equals the conditional expected value of $f$ relative to $\mu$ and $\Pi_i$, i.e., $E_i^{t_i}(f \mid \Pi_i(\omega)) = E_i^\mu(f \mid \Pi_i(\omega))$.

## 3.2 Delusional belief revision

Now replace $\Pi_i$ by $b_i$ in Equations (1) and (3).

Let $\mu$ be a probability distribution over $\Omega$, and let $b_i$ be a belief structure over $\Omega$ with corresponding partition $\Pi_i$. We introduce here the *delusional revision* of $\mu$ at $\omega$ according to $b_i$, defining it as the probability distribution $\widehat{\mu}(\omega)$ such that

$$\widehat{\mu}(\omega)(\omega') = \begin{cases} \frac{\mu(\omega')}{\mu(b_i(\omega))} & \text{if } \omega' \in b_i(\omega) \\ 0 & \text{otherwise} \end{cases} \quad (4)$$

if $\mu(b_i(\omega)) > 0$; otherwise it is undefined.

Let $f$ be a random variable over $\Omega$, let $\mu$ be a probability distribution over $\Omega$, and let $b_i$ be a belief structure over $\Omega$ with corresponding partition $\Pi_i$. Then the *delusional conditional expected value of $f$ at $\omega$ according to $b_i$* is

$$E_i^\mu(f \mid b_i(\omega)) := \frac{1}{\mu(b_i(\omega))} \sum_{\omega' \in b_i(\omega)} f(\omega')\mu(\omega)(\omega'), \quad (5)$$

if $\mu(\Pi_i(\omega)) \neq 0$ (otherwise it is not defined).

Let $(t_i)_{i \in I}$ be a probabilistic belief structure over $\Omega$, with $(\Pi_i)_{i \in I}$ the corresponding partition. Let $b_i$ be the belief structure induced by $t_i$. A *delusional prior* for $t_i$ is a probability distribution $\mu \in \Delta(\Omega)$, such that $\widehat{\mu}(\omega) = t_i(\omega)$ at each $\omega$, where $\widehat{\mu}(\omega)$ is the delusional revision of $\mu$ at $\omega$ according to $b_i$ as defined in Equation (4). A *common delusional prior* for $(t_i)_{i \in I}$ is a probability distribution $\mu \in \Delta(\Omega)$ that is a prior for each $t_i$.

Let $\phi_i$ be a standard prior for $t_i$, and suppose that for a state $\omega$, $t_i(\omega)(\omega) = 0$, and therefore that $\omega \in \Pi_i(\omega)$ but $\omega \notin b_i(\omega)$. Then by Equation (1) it must be the case that $\phi_i(\omega) = 0$. The same reasoning does not hold for a delusional prior; a standard prior is a delusional prior, but the converse is not necessarily true.

EXAMPLE 1. *Consider a one-player probabilistic belief structure over a state space $\Omega = \{\omega_1, \omega_2, \omega_3\}$ defined by*

$$t(\omega_k)(\omega_1) = 0; \; t(\omega_k)(\omega_2) = \frac{1}{2}; \; t(\omega_k)(\omega_3) = \frac{1}{2}$$

*for $k \in \{1, 2, 3\}$:*

$$t = \boxed{\overbrace{\omega_1}^{0} \; \overbrace{\omega_2}^{\frac{1}{2}} \; \overbrace{\omega_3}^{\frac{1}{2}}}.$$

*This induces a belief structure*

$$b(\omega_1) = b(\omega_2) = b(\omega_3) = \{\omega_2, \omega_3\},$$

*with $\omega_1$ a deluded state, visualised as*

$$\boxed{\omega_1} \; \boxed{\omega_2 \;\; \omega_3}$$

*The probability structure has only one (standard) prior, $\mu = (0, 1/2, 1/2)$, but it has an infinite number of delusional priors. The set of delusional priors includes, for example, $(0, 1/2, 1/2)$ and $(1/3, 1/3, 1/3)$.* ♦

## 3.3 Interpersonal Belief Credibility

S5 knowledge structures, by dint of satisfying the truth axiom, satisfy the property that $\bigcap_{i \in I} b_i(\omega) \neq \emptyset$ for all states $\omega \in \Omega$.

In KD45 belief structures there may be states at which $\bigcap_{i \in I} b_i(\omega) = \emptyset$. When

$$\bigcap_{i \in I} b_i(\omega) \neq \emptyset$$

for all states $\omega$ we will say that the belief structure satisfies *interpersonal belief credibility*.

## 4. COMMON BELIEF

Denote $b(\omega) = \bigcup_{i \in I} b_i(\omega)$ and let $b^m$ be the composition of the function $k$ repeated $m$ times. Furthermore, define for each $\omega$ the *common belief set* $b^Q(\omega)$ of $\omega$ in $\Omega$ by

$$b^Q(\omega) := \bigcup_{m \geq 1} b^m(\omega) \quad (6)$$

S5 knowledge structures are naturally partitioned into common knowledge components. Let $\{\Omega, (k_i)_{i \in I}\}$ be a knowledge structure. The *meet* is the finest common coarsening of the players' partitions. Each element of the meet of $\mathbf{\Pi}$ is called a *common knowledge component* of $\mathbf{\Pi}$. Denote by $C(\omega)$ the common knowledge component of a state $\omega$ in a knowledge structure.

Let $T \subseteq \Omega$ be a common knowledge component. $T$ can be characterised in several ways. One way is by knowledge chains. Defining $k : \Omega \to 2^{\Omega}$ by $k(\omega) := \bigcup_{i \in I} \Pi_i(\omega)$ and for $m \geq 0$ letting $k^m$ be the composition of the function $k$ repeated $m$ times, it is well known that $T = \bigcup_{m \geq 1} k^m(\omega)$ for any $\omega \in T$.

In addition, in S5 knowledge structures, a common knowledge component at $\omega$ can be characterised by the fact that

$$C(\omega) = \bigcup_{\omega \in T} \Pi_i(\omega).$$

for *all* players $i \in I$

The corresponding statement in KD45 does not hold, i.e., it is not always the case that $b^Q(\omega) = \bigcup_{\omega' \in \Omega_0} b_i(\omega')$ for some $\omega_0 \subseteq \Omega$. When it does we will want to take note of this.

*Definition 2.* There is strong common belief in truth at a state $\omega$ if there exists $\Omega_0 \subseteq \Omega$ such that $b^Q(\omega) = \bigcup_{\omega' \in \Omega_0} b_i(\omega')$ for all $i \in I$. ♦

PROPOSITION 1. *There is strong common belief in truth at every state iff the belief structure is non-singular.*

## 5. AGREEMENT IN BELIEF STRUCTURES

### 5.1 Standard No Betting

*Definition 3.* An $n$-tuple of random variables $\{f_1, \ldots, f_n\}$ is a *bet* if $\sum_{i=1}^n f_i = 0$. ♦

*Definition 4.* Let $(t_i)_{i \in I}$ be a probabilistic belief structure. Then a bet is an *agreeable bet* at $\omega$ (relative to $(t_i)$) if $E_i^{t_i}(f \mid \Pi_i(\omega)) > 0$ for all $i \in I$. A bet $f$ is a *common knowledge agreeable bet* at $\omega$ if it is common knowledge at $\omega$ that $f$ is an agreeable bet. ♦

The main characterisation of the existence of common priors in S5 knowledge models in the literature is what is sometimes known as the No Betting Theorem: a finite type space has a common prior if and only if there does not exist a common knowledge agreeable bet at any $\omega$. In the special case of a two-player probabilistic belief structure where the random variable is the characteristic function

$$1^H(\omega) = \begin{cases} 1 & \text{if } \omega \in H \\ 0 & \text{if } \omega \notin H \end{cases}$$

where $H$ is an event, this characterisation implies the seminal Aumann Agreement Theorem ([Aumann (1976)]), which states that if it is common knowledge at a state of the world that player 1 ascribes probability $\eta_1$ to event $H$ and player 2 ascribes probability $\eta_2$ to the same event, then $\eta_1 = \eta_2$.

### 5.2 KD45 No Betting

*Definition 5.* Let $(t_i)_{i \in I}$ be a probabilistic belief structure and $(b_i)_{i \in I}$ a belief structure induced by $(t_i)_{i \in I}$. A bet $f$ is a *common belief agreeable bet at $\omega$* if it is common belief at $\omega$ that $f$ is an agreeable bet. ♦

With these definitions, we can now ask whether an analogue to the No Betting Theorem of S5 models holds in the KD45 setting. Given a probabilistic belief structure $(t_i)_{i \in I}$, does the existence of a common delusional prior imply that there is no common belief agreeable bet?

The answer to this question is no, as the following example[4] shows.

EXAMPLE 2. *Let $\Omega = \{\omega_1, \omega_2, \omega_3\}$. Consider the two-player probabilistic belief structure $(t_1, t_2)$ defined by*

$$t_1(\omega_k)(\omega_1) = \frac{1}{3}; t_1(\omega_k)(\omega_2) = \frac{1}{3}; t_1(\omega_k)(\omega_3) = \frac{1}{3},$$

*and*

$$t_2(\omega_k)(\omega_1) = 0; t_2(\omega_k)(\omega_2) = \frac{1}{2}; t_2(\omega_k)(\omega_3) = \frac{1}{2}$$

*for $k \in \{1, 2, 3\}$:*

$$t_1 = \overbrace{\underbrace{\omega_1}_{}\underbrace{\omega_2}_{}\underbrace{\omega_3}_{}}^{\frac{1}{3} \quad \frac{1}{3} \quad \frac{1}{3}},$$

$$t_2 = \overbrace{\underbrace{\omega_1}_{}\underbrace{\omega_2}_{}\underbrace{\omega_3}_{}}^{0 \quad \frac{1}{2} \quad \frac{1}{2}}.$$

*This induces the belief structure $(b_1, b_2)$*

$$b_1(\omega_1) = b_1(\omega_2) = b_1(\omega_3) = \{\omega_1, \omega_2, \omega_3\},$$

*and*

$$b_2(\omega_1) = b_2(\omega_2) = b_2(\omega_3) = \{\omega_2, \omega_3\},$$

*visualised as*

$$\boxed{\omega_1 \quad \omega_2 \quad \omega_3}$$

$$\omega_1 \quad \boxed{\omega_2 \quad \omega_3}$$

*For this belief structure, $\mu = (1/3, 1/3, 1/3)$ is a common delusional prior. Let $H = \{\omega_1, \omega_2\}$. Then it is common belief at every state $\omega$ that $E_1^{t_1}(1^H \mid b_1(\omega)) = 2/3$, while $E_2^{t_2}(1^H \mid b_2(\omega)) = 1/2$.* ♦

To recapitulate something resembling the No Betting Theorem in belief structures, we add a new definition.

*Definition 6.* There is *weak common belief in truth*[5] at a state $\omega$ if there exists a state $\omega' \in b^Q(\omega)$ at which there is strong common belief in truth. ♦

An equivalent way of stating the content of Definition 6 is as follows: there is weak common belief in truth at $\omega$ iff there exists a state $\omega' \in b^Q(\omega)$ such that

$$\bigcup_{\omega'' \in b^Q(\omega')} b_i(\omega'') = \bigcup_{\omega'' \in b^Q(\omega')} b_j(\omega'')$$

for all $i, j \in I$. This can be read intuitively as the players 'eventually' getting to strong common belief in truth as they follow chains in the common belief set.

A belief structure version of the No Betting Theorem can be attained if we assume weak common belief in truth.

---

[4] This example is inspired by an example in [Collins (1997)].
[5] Although weak common belief in truth may seem abstract at first reading, it arises naturally in the study of interactive belief models. Concepts very similar to that of weak common belief in truth are introduced and used in [Battigalli and Bonanno (1999)] and [Tarbush (2011)].

THEOREM 1. *Let $(t_i)_{i \in I}$ be a probabilistic belief structure over $\Omega$ and let $\omega$ be a state at which there is weak common belief in truth. Then there is a common delusional prior if and only if there is no common belief agreeable bet at $\omega$.*

Since strong common belief in truth implies weak common belief in truth, and in a non-singular probabilistic belief structure there is strong common belief in truth at every state, Theorem 2 (which is close in content to a result appearing in [Bonanno and Nehring (1999)]) follows from Theorem 1 as a corollary.

THEOREM 2. *Let $(t_i)_{i \in I}$ be a non-singular probabilistic belief structure over $\Omega$. Then there is a common delusional prior if and only if there is no common belief agreeable bet at any state $\omega \in \Omega$.*

EXAMPLE 3. *The state space consists of $\{0, 1, 2, 3, 4, 5, 6, 7\}$. There are two players, $i$ and $j$. The belief structure*

$$((\Pi_i, b_i), (\Pi_j, b_j))$$

*is as follows:*
*Player $i$'s beliefs are*

| 1 | 2 | 3 | 4 | 5 | 6 | 7 |

*Player $j$'s beliefs are*

| 1 | 2 | 3 | 4 | 5 | 6 | 7 |

*The states 3 and 4 are delusional states for both player $i$ and player $j$, hence they perceive the same world. Note also that $b_i(3) = \{5\}$ while $b_j(3) = \{1, 2\}$, and this structure therefore does not satisfy interpersonal belief credibility. In fact, the structure can naturally be divided into two 'certainty components', $\{1, 2\}$ and $\{5, 6, 7\}$; at states 3 and 4, player $i$ is certain that the true component is $\{5, 6, 7\}$ while player $j$ is certain that the true component is $\{1, 2\}$.*

*The above belief structure can be induced by the following non-singular probabilistic belief structure $(t_i, t_j)$:*

$$t_i = \begin{array}{|cc|cc|c|cc|} \hline 1 & 1 & 0 & 0 & 1 & 1/2 & 1/2 \\ 1 & 2 & 3 & 4 & 5 & 6 & 7 \\ \hline \end{array}$$

$$t_j = \begin{array}{|cccc|ccc|} \hline 1/2 & 1/2 & 0 & 0 & 1/3 & 1/3 & 1/3 \\ 1 & 2 & 3 & 4 & 5 & 6 & 7 \\ \hline \end{array}$$

*This probabilistic belief structure has an infinite number of common delusional priors; for example,*

$$\mu = (\frac{1}{7}, \frac{1}{7}, \frac{1}{14}, \frac{1}{14}, \frac{1}{7}, \frac{1}{7}, \frac{1}{7}).$$

*There can therefore be no common belief disagreement.*

*We close by noting the following. Suppose that are working in the standard S5 knowledge model (hence that the players make 'no mistakes', that is, they revise beliefs perfectly correctly), and that the players start out with two separate priors, given by*

$$\mu_i = (\frac{1}{7}, \frac{1}{7}, \frac{1}{28}, \frac{3}{28}, \frac{1}{7}, \frac{1}{7}, \frac{1}{7})$$

*and*

$$\mu_j = (\frac{1}{7}, \frac{1}{7}, \frac{1}{14}, \frac{1}{14}, \frac{1}{7}, \frac{1}{7}, \frac{1}{7}).$$

*Then the players will revise their beliefs into the following posteriors*

$$\widehat{t_i} = \begin{array}{|c|c|cc|c|cc|} \hline 1 & 1 & 1/8 & 3/8 & 1/2 & 1/2 & 1/2 \\ 1 & 2 & 3 & 4 & 5 & 6 & 7 \\ \hline \end{array}$$

$$\widehat{t_j} = \begin{array}{|cccc|ccc|} \hline 1/3 & 1/3 & 1/6 & 1/6 & 1/3 & 1/3 & 1/3 \\ 1 & 2 & 3 & 4 & 5 & 6 & 7 \\ \hline \end{array}.$$

*Defining a bet $(f_i, -f_i)$ by*

$$f_i = (1/4, 1/4, -6, 3, -1/8, 1/32, 1/32),$$

*it can be checked that this bet is common knowledge agreeable at every state. But if the players make mistakes, using delusional revision with both players having deluded states at 3 and 4, then instead of $\widehat{t_i}$ and $\widehat{t_j}$ they will derive the posteriors $t_i$ and $t_j$, which as we have seen have a common delusional prior precluding disagreement.* ♦

## 6. REFERENCES

[Aumann (1976)] Aumann, R. J. (1976), Agreeing to Disagree, *Ann. Statist.*, 4(6), 1236–1239.

[Battigalli and Bonanno (1999)] Battigalli, P., and G. Bonanno (1999), Recent Results on Belief, Knowledge and the Epistemic Foundations of Game Theory *Review of Economic Studies*, 64, 23–46.

[Bonanno and Nehring (1999)] Bonanno, G., and K. Nehring (1999), How to Make Sense of the Common Prior Assumption under Incomplete Information, *Int. J. Game Theory* 28, 409–434.

[Collins (1997)] Collins, J., (1997), How We can Agree to Disagree. *Working paper.*

[Geanakoplos (1989)] Geanakoplos, J. (1989), Game Theory Without Partitions and Applications to Speculation and Consensus, *Cowles Foundation Discussion Paper 914.*

[Geanakoplos and Polemarchakis (1982)] Geanakoplos, J., and H. Polemarchakis (1982), We Can't Disagree Forever, *Journal of Economic Theory*, (28), 192–200.

[Hart and Tauman (2004)] Hart, S., and Y. Tauman (2004), Market Crashes without External Shocks, *Journal of Business*, (77), 1–8.

[Morris (1996)] Morris, S. (1996), The Logic of Belief and Belief Change: A Decision Theoretic Approach, *Journal of the Economic Theory*, (69), 1–23.

[Samet (2011)] Samet, D. (2011), Common Belief of Rationality in Games of Perfect Information, *Working Paper.*

[Tarbush (2011)] Tarbush, B. (2011), Agreeing to disagree with generalised decision functions, *University of Oxford, Department of Economics, Working Paper.*

## APPENDIX

**Proof of Theorem 1.** We first add a definition and a lemma, for the sake of proving the theorem.

*Definition 7.* Let $(t_i)_{i \in I}$ be a probabilistic belief structure over $\Omega$ with corresponding partition profile $\Pi := (\Pi_i)_{i \in I}$, and let $X \subset \Omega$ be a subset of $\Omega$. Define $\Pi$ *restricted to* $X$, denoted $\Pi^X$, to be the partition profile over $X$ given by $\Pi_i^X(\omega) := \Pi_i(\omega) \cap X$ for any state $\omega$. Further, for each $i \in I$ let $t_i^X$ be any type function over $(X, \mathbf{\Pi}^X)$ that satisfies the property that for any $\omega \in \Omega$, $t_i(\omega)(\Pi_i^X)t_i^X(\omega) = t_i(\omega)$. ♦

Intuitively, $\Pi_i^X$ is the partition of $X$ derived from the partition $\Pi_i$ of $\Omega$ by 'ignoring all states outside of $X$'. It then follows intuitively that $t_i^X(\omega)$, for each state $\omega \in X$, is $t_i(\omega)$ scaled relative to the other states in $\Pi_i^X(\omega)$ in such a way that $\sum_{\omega \in X} t_i^X(\omega) = 1$.

For a random variable $f$, denote

$$E_i^X(f \mid \Pi_i^X(\omega)) := \sum_{\omega' \in \Pi_i^X(\omega)} t_i^X(\omega') f(\omega').$$

A bet $\{f_1, \ldots, f_n\}$ is an *agreeable bet relative to* $(t_i^X)_i$ at $\omega \in X$ if $E_i^X(f \mid \omega) > 0$ for all $i \in I$. We will say that it is simply an agreeable relative to $(t_i^X)_i$ if it is an agreeable bet relative to $(t_i^X)_i$ at all states $\omega \in X$.

LEMMA 1. *Let $(t_i)_{i \in I}$ be a probabilistic belief structure over $\Omega$, let $\omega \in \Omega$ and let $X$ be a non-empty subset of $b^Q(\omega)$, the common belief set of $\omega$. Suppose that there exists an agreeable bet relative to $(t_i^X)_i$. Then there exists an agreeable bet relative to $b^Q(\omega)$.*

**Proof.** Let $f$ be an agreeable bet relative to $(t_i^X)_i$. If $X = b^Q(\omega)$, there is nothing to prove.

Otherwise, we distinguish a few cases:

1. Suppose that there exists a state $\omega'' \in X$ such that $b_i(\omega'') \setminus X \neq \emptyset$ for some $i \in I$. Let $\omega' \in b_i(\omega'') \setminus X$ (hence $t_i(\omega') > 0$), and let $\varepsilon := E_i^X(f_i \mid \Pi_i^X(\omega')) = E_i^X(f_i \mid \Pi_i^X(\omega''))$. By assumption, $\varepsilon > 0$ (since $f$ is an agreeable bet relative to $(t_i^X)_i$). Set $Y := X \cup \omega'$.

   Next, let $\overline{f}_i(\omega')$ be a negative real number satisfying

   $$0 > \overline{f}_i(\omega') > \frac{-(1-t_i^Y(\omega'))}{t_i^Y(\omega')} \varepsilon,$$

   and for $j \neq i$, set $\overline{f}_j(\omega') := -\overline{f}_i(\omega')/(n-1) > 0$, where $n = |I|$.

   Clearly, by construction, $\sum_{j \in I} \overline{f}_j(\omega') = 0$. Complete the definition of $\overline{f}$ by letting $\overline{f}(\omega''') := f(\omega''')$ for all $\omega''' \in X$. It is straightforward to check that $\overline{f}$ is an agreeable bet relative to $(t_i^Y)_{i \in I}$.

2. Suppose that there is a state $\omega' \in b^Q(\omega) \setminus X$ such that $b_i(\omega') \cap X \neq \emptyset$. Set $Y := X \cup \omega'$.

   We distinguish two sub-cases:

   (a) If $t_i(\omega') = 0$, then for all $j \in I \setminus i$ let $\overline{f}_j(\omega')$ be any arbitrary positive number, and set $\overline{f}_i(\omega') = -\sum_{j \in I \setminus i} \overline{f}_j(\omega')$. Then $\overline{f}$ is an agreeable bet relative to $(t_i^Y)_{i \in I}$.

   (b) If $t_i(\omega') > 0$, let $\varepsilon := E_i^X(f_i \mid \Pi_i^X(\omega'))$. By assumption, $\varepsilon > 0$ (since $b_i(\omega') \cap X \neq \emptyset$ and $f$ is an agreeable bet relative to $(t_i^X)_i$). From this point, define $\overline{f}_j$ for all $j \in I$ exactly as in Case 1 above, yielding an agreeable bet relative to $(t_i^Y)_{i \in I}$.

Now simply repeat this procedure as often as necessary to extend the agreeable bet to every state in the finite set $b^Q(\omega)$. ∎

**Completion of the proof of Theorem 1.** Let $(t_i)_{i \in I}$ be a probabilistic belief structure over $\Omega$, and let $\omega$ be a state at which there is weak common belief in truth, and hence there is $\omega' \in b^Q(\omega)$ at which there is strong common belief in truth, i.e.,

$$\bigcup_{\omega'' \in b^Q(\omega')} b_i(\omega'') = \bigcup_{\omega'' \in b^Q(\omega')} b_j(\omega'')$$

for all $i, j \in I$. If we restrict attention solely to the states in $b^Q(\omega')$, we can consider the operators $b_i$ for all $i$ to constitute an S5 knowledge structure over $b^Q(\omega)$.

In one direction, suppose that there is a common delusional prior $\mu$. Then $\mu$ restricted to $b^Q(\omega')$ is a common (standard) prior over $b^Q(\omega')$ regarded as a knowledge structure, hence there can be no common knowledge agreeable bet at any state in $b^Q(\omega')$. If there were a common belief agreeable bet at $\omega$, then that bet would be a common knowledge agreeable bet over $b^Q(\omega')$ regarded as a knowledge structure, which we just showed cannot happen. The contradiction establishes that there is no common belief agreeable bet at $\omega$.

In the other direction, suppose that there is no common delusional prior. Then there can be no common (standard) prior over $b^Q(\omega')$ regarded as a knowledge structure, because if there were such a prior $\mu$, it could be extended to a common delusional prior $\widehat{\mu}$ over all of $b^Q(\omega)$ simply by setting

$$\widehat{\mu}(\omega'') = \begin{cases} \mu(\omega'') & \text{if } \omega'' \in b^Q(\omega') \\ 0 & \text{otherwise.} \end{cases}$$

We can then apply the standard No Betting Theorem for knowledge structures to conclude that there is a common knowledge agreeable bet $\{f_1, \ldots, f_n\}$ over $b^Q(\omega')$ as a knowledge structure, which is a common belief agreeable bet over $b^Q(\omega')$ as a belief structure. Applying Lemma 1, this can be extended to a common belief agreeable bet over all of $b^Q(\omega)$, which is what was needed to be shown. ∎